\documentclass [12pt,showpacs,showkeys]  {revtex4}
\begin{document}

\title[Short Title]{Secret sharing of quantum information via entanglement swapping in cavity QED\footnote{Foundation item: Supported by the National Science Foundation of China
under Grant (60261002).}}
\author{Ying-Qiao Zhang, Xing-Ri Jin, Shou Zhang\footnote{E-mail: szhang@ybu.edu.cn}}  \affiliation{Department
of Physics, College of Science, Yanbian University, Yanji, Jilin,
133002, China}
\begin{abstract}
We proposed a scheme on secret sharing of quantum information based
on entanglement swapping in cavity QED. In our scheme, the effects
of cavity decay and thermal field are all eliminated.
\pacs{03.67.Dd; 03.65.Ud} \keywords{Secret sharing,  Entanglement
swapping, Cavity QED}
\end{abstract}
\maketitle
 \section{Introduction}
 Using the theory of quantum mechanics in the field of information
 in the recent years has produced many interesting developments,
 such as quantum teleportation\cite{BBCJPW93}, quantum cryptography\cite{EA91}, quantum
 secret sharing (QSS)\cite{HBB99}, and so on.
Quantum secret sharing, firstly proposed by Hillery \emph{et
al}\cite{HBB99}, is one of the basilic components of quantum
communication and is used to fulfill the task of classical secret
sharing. The basic idea of secret sharing, invented by both
Shamir\cite{AS79} and Blakely\cite{GRB79} independently in 1979,
is to distribute a secret between $n$ players in such a way that
any group of $k$ or more players can together reconstruct the
secret but no group of less than $k$ players can know anything
about the secret even if they cooperate. Such a system is called a
$(k, n)$-threshold scheme. The property of QSS, being used to
share both classical information and quantum information, makes it
differ from the classical secret sharing. QSS scheme can be used
in joint sharing of quantum money\cite{SW83}, sharing
difficult-to-construct ancilla states\cite{GC99}, and so on.

Therefore, many attentions\cite{CGL99, DG99, HFC03, BK03, SG01a,
LSBSL04, KKI99, SB00, KBB02, GG03, XLDP04, DLWX04, TZG01, DLLZW05}
have been concentrated on the realization of QSS both in theory
and experiment using many kinds of methods. Entanglement swapping
is one of the methods being used in QSS protocols\cite{KBB02,
MZ04, LZP04}. Zhang \emph{et al}\cite{MZ04} and Li \emph{et
al}\cite{LZP04} proposed multi-party quantum secret sharing
protocols based on entanglement swapping using the Bell-state
measurements and GHZ-basis measurements, respectively. But
performing the Bell-state measurements and GHZ-basis measurements
in QSS scheme is difficult, so we propose a scheme to realize the
QSS in cavity QED via entanglement swapping. We know that a $(k,
n)$-threshold scheme requires that no single player can have any
information on the unknown state if they act alone. So our scheme
isn't the conventional quantum $(k, n)$-threshold scheme, because
each player has the amplitude information of the unknown state. In
Section II, we discuss the secret sharing protocol in three-party
system, and the generalization to multi-party system is given in
Section III.

\section{Three-party quantum information secret sharing}
We consider the three-party system consists of Alice, Bob, and
Charlie. At first, Alice possesses six atoms, namely, atom 1, atom
2,..., atom 6. The state of atom 1 that Alice wants to send to Bob
and Charlie is
\begin{equation}\label{e01}
|\Psi\rangle_{1}=\alpha|e\rangle_{1}+\beta|g\rangle_{1},
\end{equation}
where $\alpha$ and $\beta$ are unknown coefficients, they satisfy
$|\alpha|^{2}+|\beta|^{2}$=1. $|e\rangle$ and $|g\rangle$ are atom
excited and ground states, respectively. The states of atoms 2, 3,
4, and 5, 6 are in three-atom maximally entangled state and
two-atom maximally entangled state, respectively, as
\begin{equation}\label{e02}
|\Psi\rangle_{234}=\frac{1}{\sqrt{2}}(|eee\rangle_{234}+|ggg\rangle_{234}),
\end{equation}
\begin{equation}\label{e03}
|\Psi\rangle_{56}=\frac{1}{\sqrt{2}}(|ee\rangle_{56}+|gg\rangle_{56}).
\end{equation}
We consider the atoms 1, 2, 3, 4, 5, and 6 are all identical
two-level atoms. The joint state of the six atoms can be expressed
as
\begin{equation}\label{e04}
|\Psi\rangle=|\Psi\rangle_{1}\otimes|\Psi\rangle_{234}\otimes|\Psi\rangle_{56}.
\end{equation}

Firstly, Alice introduces two identical single-mode cavities and
simultaneously sends the atoms 1, 2 and atoms 3, 5 into the two
single-mode cavities, respectively. So there are two interaction
systems of atoms and cavity in all. Considering the two atoms 1, 2
(3, 5) simultaneously interacting with the single-mode cavity
field and driving by the classical field, respectively.

The interaction Hamiltonian between the atoms and the single-mode
cavity is\cite{YC04} ($\hbar=1$)
\begin{equation}\label{e05}
H=\omega_{0}\sum\limits_{j=1}^{2}S_{z,j}+\omega_{1}a^{\dag}a+
\sum\limits_{j=1}^{2}[g(a^{\dag}S_{j}^{-}+aS_{j}^{\dag})+
\Omega(S_{j}^{\dag}e^{-i\omega_{2}t}+S_{j}^{-}e^{i\omega_{2}t})],
\end{equation}
where $S_{j}^{-}=|g\rangle_{jj}\langle e|$,
$S_{j}^{\dag}=|e\rangle_{jj}\langle g|$,
$S_{z,j}=\frac{1}{2}(|e\rangle_{jj}\langle
e|-|g\rangle_{jj}\langle g|)$, $|e\rangle_{j}$ and $|g\rangle_{j}$
are the excited and ground states of the $j$th atom,  $a^{\dag}$
and $a$ are creation operator and annihilation operator of the
cavity mode. $g$ is the coupling constant between the atoms and
cavity, $\omega_{0}$, $\omega_{1}$, $\omega_{2}$ are atomic
transition frequency $(e\leftrightarrow g)$, cavity frequency,
driving field frequency, respectively, and $\Omega$ is the Rabi
frequency of the classical field. We consider the atomic
transition frequency equals to driving field frequency
($\omega_{0}=\omega_{2}$). In the case of large detuning
$\delta\gg g/2$ and strong driving field $2\Omega\gg\delta$ ($g$
limit), the effective Hamiltonian of the interaction system can be
expressed as\cite{SBZ04}
\begin{equation}\label{e06}
H_{\rm eff}=\frac{\lambda}{2}[\sum\limits_{j=1}^{2}
(|e\rangle_{jj}\langle e|+|g\rangle_{jj}\langle
g|)+\sum\limits_{j,k=1,j\neq k}^{2}
(S_{j}^{\dag}S_{k}^{\dag}+S_{j}^{\dag}S_{k}^{-}+\texttt{H.c.})],
\end{equation}
where $\lambda= g^{2}/2\delta$ with $\delta$ being the detuning
between $\omega_{0}$ and $\omega_{1}$. So the effects of cavity
decay and thermal field are all avoided. The evolution operator of
the system in interaction picture can be expressed as
\begin{equation}\label{e07}
U(t)=e^{-iH_{0}t}e^{-iH_{\rm eff}t},
\end{equation}
where $H_{0}=\sum\limits_{j=1}^{2}\Omega(S_{j}^{\dag}+S_{j}^{-})$.

 We consider the interaction time of atoms 1, 2 with the single-mode cavity
 and the interaction time of atoms 3, 5 with the single-mode cavity
 are the same. After the interaction,
Alice sends atoms 4 and 6 to Bob and Charlie, respectively. When
she is sure that Bob and Charlie have both receive an atom, she
measures on atoms 1, 2, 3, 5 and informs Bob and Charlie of her
measurement results via a public channel. If the measurement
result is $|eeee\rangle_{1235}$, the state of atoms 4, 6 collapses
into
\begin{equation}\label{e10}
|\Psi\rangle_{46}=\alpha |ee\rangle_{46}-\beta |gg\rangle_{46},
\end{equation}
by selecting the interaction time satisfy $\lambda
t=\frac{\pi}{4}$ and making the Rabi frequency satisfy $\Omega
t=\pi$. Here we must emphasize that the net effect of the
evolution is to apply a $\sigma_{z}$ to the unknown state and then
to apply a CNOT to the unknown state and a standard $|g\rangle$,
namely
$\sigma_{z}(\alpha|e\rangle+\beta|g\rangle)=\alpha|e\rangle-\beta|g\rangle$,
$\rm
CNOT$$\rightarrow(\alpha|e\rangle-\beta|g\rangle)|g\rangle=\alpha|ee\rangle-\beta|gg\rangle$.

 Now Alice has successfully transferred the quantum information
  to Bob and Charlie by entanglement swapping, so the distribution of quantum information is
 completed.

 We observe that neither Bob nor Charlie can
 recover the state $|\Psi\rangle_{1}$ in its exact form by performing any general operations themselves
  without communicating between themselves.
 Though they have the amplitude information, that is not sufficient
 since the phase information is not available. In this case they must agree to cooperate
  among themselves. Only by this way, one of them, not both, can recover the desired state
  for the no-cloning theorem.

We rewrite the state
 $|\Psi\rangle_{46}$ in Eq.(\ref{e10}), as
\begin{equation}\label{e11}
|\Psi\rangle_{46}=\frac{1}{\sqrt{2}}
[\frac{1}{\sqrt{2}}(|e\rangle_{4}+|g\rangle_{4})(\alpha|e\rangle_{6}-\beta|g\rangle_{6})+
\frac{1}{\sqrt{2}}(|e\rangle_{4}-|g\rangle_{4})(\alpha|e\rangle_{6}+\beta|g\rangle_{6})].
\end{equation}
If Alice assigns Charlie to recover the quantum state in
Eq.(\ref{e01}), then Bob needs to measure on atom 4 in the
$X$-basis, where the $X$-eigenstates are defined by
\begin{equation}\label{e12}
|X^{\pm}\rangle=\frac{1}{\sqrt{2}}(|e\rangle\pm|g\rangle).
\end{equation}
If Bob's measurement result is
\begin{equation}\label{e13}
|\Psi\rangle_{4}=\frac{1}{\sqrt{2}}(|e\rangle_{4}+|g\rangle_{4}),
\end{equation}
according to Eq.(\ref{e11}), the state of atom 6 becomes
\begin{equation}\label{e14}
|\Psi\rangle_{6}=\alpha|e\rangle_{6}-\beta|g\rangle_{6}.
\end{equation}
Then Bob communicates his outcome to Charlie in a public channel.
At this stage, Charlie can recover the unknown state by performing
the rotation operation $\sigma_{z}$ on atom 6
\begin{equation}\label{e15}
\sigma_{z}|\Psi\rangle_{6}=\alpha|e\rangle_{6}+\beta|g\rangle_{6}.
\end{equation}
So Charlie can recover the state $|\Psi\rangle_{1}$ with the help
of Bob.

Now we discuss the security of our scheme. Suppose that there is
an adversary, he will be either Bob or Charlie. The adversary
(say, Bob) wants to eavesdrop Alice's information without being
detected. If Alice assigns Bob to receive the state and Charlie
agrees to cooperate with Bob, Bob can eavesdrop the state with a
successful probability of 100 percent, and the cheating will not
be detected. If Alice assigns Bob to receive the state and Charlie
doesn't agree to cooperate with Bob, in this case Charlie doesn't
tell Bob his measurement results. Bob can also eavesdrop the
state, but the successful probability is only 50 percent, he will
still get nothing with a probability of 50 percent. However, if
Alice assigns Charlie to receive the state, Bob will measure the
state of his atom in the $X$ basis and tell his measurement
results to Charlie, Charlie can recover the state with the help of
Bob. There is also the probability that Bob could lie about his
measurement results. By doing so, Bob gains nothing and Charlie
can't recover the correct state. Of course, Bob can also manage to
get a hold of the atom that Alice sends to Charlie, and sends
Charlie an atom that he has prepared. He wants to discover the
state of Alice's atom 1 without the help of Charlie. In this way,
only when Alice assigns him to recover the state, he can get the
state of Alice's atom 1 without being detected. On the other hand,
if Alice assigns Charlie to recover the state, then Bob has some
trouble. Bob doesn't know Alice's measurement result and therefore
the atom that he sends to Charlie is not in the correct quantum
state. So the state recovered by Charlie will be different with
the state Alice has sent. If Alice checks a subset of the state
with Charlie publicly, the eavesdropping behavior can be revealed.
The security of the present scheme is the same as that in
contribution\cite{LZP04}.
\section{Multi-party quantum information secret sharing}
In this section, we generalize the three-party secret sharing
scheme to multi-party system.
 At first, Alice possesses $3n$ identical two-level atoms, marked as: atom 1, atom 2,..., atom $3n$. The state of
atom 1 that Alice wants to send to $n$ users is still in
Eq.(\ref{e01}). The states of atoms 2, 3,..., $(3n-2)$ are in the
following $(n-1)$ three-atom maximally entangled states,
respectively, as
\begin{eqnarray}\label{e16}
&&|\Psi\rangle_{234}=\frac{1}{\sqrt{2}}(|eee\rangle_{234}+|ggg\rangle_{234}),\cr\cr&&
|\Psi\rangle_{567}=\frac{1}{\sqrt{2}}(|eee\rangle_{567}+|ggg\rangle_{567}),\cr\cr&&
|\Psi\rangle_{8,9,10}=\frac{1}{\sqrt{2}}(|eee\rangle_{8,9,10}+|ggg\rangle_{8,9,10}),\cr\cr&&......,\cr\cr&&
|\Psi\rangle_{3n-4,3n-3,3n-2}=\frac{1}{\sqrt{2}}(|eee\rangle_{3n-4,3n-3,3n-2}+|ggg\rangle_{3n-4,3n-3,3n-2}).
\end{eqnarray}
The state of atoms $(3n-1)$ and $(3n)$ is in the two-atom
maximally entangled state
\begin{equation}\label{e17}
|\Psi\rangle_{3n-1,3n}=\frac{1}{\sqrt{2}}(|ee\rangle_{3n-1,3n}+|gg\rangle_{3n-1,3n}).
\end{equation}
Firstly Alice simultaneously puts $n$ pairs of atoms, namely,
atoms (1, 2), atoms (3, 5), atoms (6, 8),..., atoms $(3n-3,
3n-1)$, into $n$ identical single-mode cavities, respectively.
 So
there are $n$ interaction systems of atoms and cavity in all.
Alice selects the same interaction time $t$ for the $n$
interaction systems. Secondly Alice sends the residual $n$ atoms,
namely, atom 4, atom 7, atom 10,..., atom $(3n-2)$, and atom
$(3n)$, to each one of the $n$ users, respectively. When she is
sure that each one of the $n$ users has received an atom, Alice
measures on the $n$ pairs of atoms that having been put into the
cavities. The $n$ users obtain a pure entangled state of the
residual $n$ atoms which contains all the information of the state
in Eq.(\ref{e01}). So the distribution of the quantum information
is completed. Then Alice publicly declares her measurement results
and assigns one user (A) to receive the state. The rest $(n-1)$
users respectively perform an $X$-basis measurement as shown in
Eq.(\ref{e12}) on their own atoms, and then inform the user (A) of
their measurement results, respectively. So the user (A) can
recover the state in Eq.(1) by performing appropriate rotation
operation on his atom according to the information he has obtained
from Alice and the rest $(n-1)$ users. The security of the
multi-party secret sharing scheme is the same as that in Section
II.

\section{Summary}

In the present scheme, the interaction system of atoms and cavity
is in large detuning and strong driving field case, the effects of
cavity decay and thermal field are all avoided, so the scheme is
feasible with the present cavity QED techniques. In addition, when
we treat the multi-party system, the $n$ pairs of atoms must be
sent simultaneously into $n$ identical single-mode cavity fields,
respectively. This will product errors between experiment
operation and theory calculation. Because of the value of $n$ is a
finite number, the effects of the errors on the fidelity of the
result state can be neglected.

In conclusion, we have presented a protocol of three-party quantum
information secret sharing via entanglement swapping in cavity
QED. Our scheme is easier to realize for without performing any
Bell-state measurements and GHZ-basis measurements. This scheme
can also be generalized to the multi-party system.

\end{document}